\documentclass[aps,pra,superscriptaddress]{revtex4}
\pdfoutput=1
\usepackage[intlimits]{amsmath}
\usepackage{amsfonts}
\usepackage{psfrag}
\usepackage{enumerate}
\usepackage{subfigure}
\usepackage{enumitem}
\usepackage[usenames]{color}
\advance\voffset by -1cm
\advance\hoffset by -0.5cm
\newcommand{\beq}{\begin{eqnarray}}
\newcommand{\eeq}{\end{eqnarray}}

\newcommand\doi[2]        {\href{http://dx.doi.org/#1}{#2}}

\usepackage{epsfig}
\usepackage{color}
\usepackage{psfrag}
\usepackage{epstopdf}
\usepackage{amsmath}
\usepackage{graphicx}
\usepackage{amsfonts}
\usepackage{amssymb}
\usepackage{tikz}
\begin{document}
\bibliographystyle{plainnat}

\title{{\Large {\bf Planar quantum quenches:\\ 
Computation of exact time-dependent correlation functions at large $N$}}}

\author{Axel Cort\'es Cubero}
\affiliation{SISSA and INFN, Sezione di Trieste, via Bonomea 265, I-34136, 
Trieste, Italy}
\email{acortes@sissa.it}

\begin{abstract}
\noindent
 
We study a  quantum quench of an integrable quantum field theory in the planar infinite-$N$ limit. Unlike 
isovector-valued $O(N)$ models, matrix-valued field theories in the infinite-$N$ limit are not solvable by the Hartre-Fock approximation, and are nontrivial interacting theories. We study quenches with initial states that are color-charge neutral, correspond to integrability-preserving boundary conditions,  and that lead to nontrivial correlation functions of operators.   We compute exactly at infinite $N$, the time-dependent one- and two-point correlation functions of the energy-momentum tensor and renormalized field operator after this quench using known exact form factors. This computation can be done fully analytically, due the simplicity of the initial state and the form factors in the planar limit. We also show that this type of quench preserves factorizability at all times, allows for particle transmission from the pre-quench state, while still having nontrivial interacting post-quench dynamics.  
\vspace{3mm}
\noindent

\end{abstract}
\maketitle


\section{Introduction}

In recent years, there has been much interest in the study of the out-of-equilibrium dynamics of isolated extended quantum systems. This interest has been fueled by new experimental techniques in condensed matter physics that have permitted the minimization of dissipation effects and any interaction of the system with the environment \cite{experiments}. This has made the study of isolated quantum systems a practical and experimentally relevant endeavor. 

A commonly studied problem is that of preparing a system at time $t=0$ in an initial state, $\vert B\rangle$, that is not an eigenstate of its Hamiltonian, $H$, then unitarily evolving this state with the new Hamiltonian, as $\vert B(t)\rangle=e^{-itH}\vert B\rangle$. The initial state is often considered to be an eigenstate (more commonly, the ground state) of some initial Hamiltonian, $H_0$. The sudden change of the Hamiltonian from $H_0$ to $H$ at $t=0$ is called a quantum quench.

One important problem is that of computing the time-dependent correlation functions of local operators, and studying their possible relaxation to a steady state value at an infinite time after the quench. This question is particularly interesting in highly symmetric systems, such as conformal or integrable theories \cite{CC,SilvaReview,IC,Berges,Rigol,FM,FM2,EMP,CEF,GGErecent,mario,MMP,nardis,spyrospasquale,ScEs,Bertin}. In these cases, the additional conservation laws prevent the system from relaxing  into a steady state described by a thermal Gibbs ensemble. It is widely believed that the relaxation of conformal and integrable models is described instead by a generalized Gibbs ensemble (GGE), which takes into account all the conserved charges (and not only the Hamiltonian, as does the Gibbs ensemble) \cite{Rigol}.

One advantage of studying integrable quantum field theories is that the extra symmetries provide powerful nonperturbative tools, which can sometimes be used to find exact analytic results. 
The first step towards an exact computation of the time evolution of observables after a quench, is to express the initial state in terms of the eigenstates of the post-quench Hamiltonian. The structure of the initial state is generally not known, even for integrable theories. However, a useful approach is to study the time evolution of some simple initial state that may or may not describe the quantum quench process, but which is analytically tractable. 

A particularly useful class of initial states, which is often referred to as an integrable initial state (we will use this nomenclature for the rest of this paper), was first discussed in Ref. \cite{GZ}.
These initial states are called ``integrable" in correspondence with the integrability-preserving boundary conditions in a model with a spatial boundary (instead of the temporal boundary in the quench problem). We point out, however, that  ``integrable initial state" is a misleading name, in that the integrability of the theory is only a property of the Hamiltonian, and consequently the time evolution. The integrability of the theory can be preserved even with a more general initial state. The particle-creation amplitudes in the integrable initial states are obtained by Wick-rotating the boundary S-matrix of the theory with a spatial boundary\footnote{ The boundary S-matrix can be obtained using the axioms proposed in \cite{GZ}. One of these axioms, the so called ``boundary unitarity axiom", is too restrictive to describe a quantum quench process, since information can be transmitted across the boundary at $t=0$, so this requirement is usually dropped when defining integrable initial states.}. These  states, which we will discuss later in more detail, can be expressed simply in terms of the post-quench particle states. 
 This expression can then be used compute observables using exact form-factors. These expansions have been used to study quenches in several integrable field theories \cite{FM,ScEs,Bertin}.

There are two general difficulties with the form factor approach to integrable quantum quenches that we will address in this paper. The first problem is that of determining  if  a proper quantum quench of an integrable theory produces an integrable initial state (or some different kind of initial state). That is,  if one prepares the system in an eigenstate of an integrable Hamiltonian, $H_0$, and suddenly changes some parameter, resulting in a (still integrable) Hamiltonian, $H$, there is no guarantee that the initial state expressed in terms of the post-quench particle states is an integrable state, like those of \cite{GZ}. 

 It has in fact been shown by Delfino \cite{Delfino} that genuine quantum quenches of integrable field theories in general break factorizability. It was shown for  scattering theories with a single particle species, that if one allows for transmission of particles from the pre-quench state accross the boundary at $t=0$, one cannot demand factorization at all times (pre- and post-quench), unless the post-quench theory is free (the S-matrix is $\pm 1$). It is then interpreted in \cite{Delfino} that this lack of factorizability implies that integrability is always broken at the $t=0$ boundary in a quench of an interacting theory. This casts even more doubt on whether it is justified to use the integrable initial states from \cite{GZ}, since it is unclear why the state should have any integrable structure. This result was later generalized by Schuricht in \cite{Schuricht} for several types of non-diagonal scattering theories with different types of particles, with the same negative result. It then seems that a fully factorizable quench  from an excited state with particles that could cross the $t=0$ boundary is impossible except for free theories.
 
 Despite these negative results of Delfino and Schurcht, it is worth mentioning that there is recent evidence that suggests the validity of the integrable initial states for a quantum quench. The quench from a free bosonic theory, into a model with a sinh-Gordon potential was studied in Ref. \cite{Spyros}. The initial state is determined by iteratively solving an infinite set of integrable equations that arise from imposing  boundary conditions for the values of the field at $t=0$. It was found that integrable states are at least a good approximation to the solution of these integral equations.

 The second difficulty with the form factor approach is that, even if one has an integrable initial state, the computation of correlation functions using form factors is still a very difficult problem. The individual terms of the form factor expansion are very difficult to compute, and in general, one can only compute the first few terms. For an integrable quench of the sine-Gordon model, in Ref. \cite{Bertin} they have been able to compute the first few terms in a direct form factor expansion (which they call the linked cluster expansion), and by extrapolating these results, deduct the behavior of the contributions from higher form factors, showing that expectation values of vertex operators decay exponentially in time into a steady-state value. This is confirmed in the same reference by a separate calculation  based on the quench action method in the low density limit, where this exponential decay is also recovered. The same exponential decay was also found from a semiclassical calculation in Ref. \cite{kormos}. A different approach has been recently proposed \cite{sinh}, based on the so-called Negro-Smirnov formula \cite{negro}, to avoid the resummation of infinite form factors, but so far, this approach has only been shown to work for computing steady-state observables in the infinite time limit.

 In this paper, we study an integrable quantum quench of a field theory in the planar large-$N$ limit. We consider a model with an $N\times N$ matrix-valued bare field, and take $N$ to infinity, while keeping the mass gap constant. The large-$N$ limit is useful because it simplifies the quench problem to a level where one can perform analytical calculations, yet keeps nontrivial interactions. In perturbation theory, the leading contributions to observables, in powers of $1/N$ come from planar Feynman diagrams (thus we call this the planar limit). This can be contrasted with the relatively trivial large-$N$ limit of the isovector-valued $O(N)$ sigma model, or $\phi^4$ models. The isovector theories in the large-$N$ limit can be solved exactly by a Hartree-Fock approximation. This solution has been known for a long time for isovector theories at zero temperature \cite{polyakov}, and has been recently also applied to the quench problem \cite{selfconsistent}. Different questions about the out-of-equilibrium dynamics, and equilibration of large-$N$ isovector theories have been studied by many authors, including \cite{vectorpapersone}, and in particular quantum quenches have been studied in \cite{vectorpaperstwo}. The Hartre-Fock approximation for the large-$N$ limit of isovector theories essentially reduces the problem to considering only a quadratic action, making the problem very similar to a free theory, and exactly solvable, even in non-equilibrium scenarios and for non-integrable theories.  Another related recent development is the study of quantum quenches in $(0+1)$-dimensional large-$N$ matrix models in \cite{matrixqm}, where the relaxation into a stationary state described by a GGE was explored.

 The precise model we will study is the principal chiral sigma model (PCSM), with the action
\beq
S_{\rm PCSM}=\int d^2 x\,\frac{1}{2g_0^2}{\rm Tr}\partial_\mu U^\dag(x)\partial^\mu U(x),\label{pcsmaction}
\eeq
where $U(x)\in {\rm SU}(N)$. This model has been shown to be integrable, and its exact S-matrix is known \cite{wiegmann}. The action (\ref{pcsmaction}) has an ${\rm SU}(N)\times {\rm SU}(N)$ global symmetry given by $U(x)\to V_L U(x) V_{R}$, with $V_{L,R}\in {\rm SU}(N)$. 

The classical PCSM is a scale invariant of massless nonlinear waves. In the quantum theory, however, the PCSM is believed to have a dynamically generated mass gap, which we call $m$. This is the only parameter in the theory (related to the bare coupling by dimensional transmutation). We will therefore be interested  in studying a sudden mass quench, where at $t=0$, the mass is changed from $m_0$ to $m$. A relation between the bare coupling constant and the mass gap can be obtained perturbatively:
\beq
m=C\Lambda \left(g_0^{K_2} e^{-\frac{K_1}{g_0^2}}+\cdots\right),\nonumber
\eeq
where $\Lambda$ is a high momentum cut-off, $K_1=-1$ and $K_2=4\pi$ are are found
from the one- and two-loop coefficients of the model's beta function \cite{mckane}, and $C$ is a non-universal constant that depends on the normalization scheme.

It is believed that at high, temperatures, the PCSM undergoes a crossover into a ``deconfined" phase, where the massless degrees of freedom of the classical theory are released. This was seen numerically in \cite{vicari} by finding a peak in the specific heat of at some critical temperature. It is also seen numerically that the peak becomes sharper as $N$ is increased, suggesting that the crossover becomes a phase transition as $N\to\infty$.

An infinite set of conserved charges for the quantum theory was found in \cite{wittencharges}. These charges can in principle be used to define a GGE. For the simple type of initial states we consider in this paper, however, as we will see later, the GGE turns out to be trivial. The long-time steady state observables seem to be given by the trivial zero-temperature expectation values. Nevertheless, it is still possible that there can exists a nontrivial GGE for more general initial states than the ones discussed here.

It is known that the large-$N$ limit of the PCSM cannot be solved by the Hartree Fock approximation \cite{polyakov}. There have been recently, however, two different analytic approaches towards an exact solution of this model. One approach (the one we continue for the rest of this paper) is to utilize the integrability of the PCSM. Using integrability, one can find the exact S-matrix \cite{wiegmann}, which was used to find exact form factors at large $N$ \cite{renormalizedfield,multiparticle,correlation}. These form factors have also been used to compute correlation functions the PCSM at finite temperature \cite{thooftthermodynamics} in the planar limit. The S-matrix has also been used to study the thermodynamics of the model through the thermodynamic Bethe ansatz \cite{kazakovleurent}. This approach from integrability assumes there is a dynamically generated mass gap, without proving this fact, and it is not based on a first-principles path integral computation. A complementary approach has been studied in Ref. \cite{unsal}, based on expanding observables in a resurgent transseries of the coupling, $g_0$, which includes information about nonperturbative complex saddle points (which exist, despite the fact that the PCSM has no topological instantons). These nonperturbative saddle points provide a mechanism for the generation of the mass gap.

The two main results of this paper are the determination of a simple class of integrable initial states  of the PCSM at infinite $N$, and the computation of exact time-dependent correlators for the corresponding quantum quench. We focus on integrable states that are color neutral, that is, we assume all the ${\rm SU}(N)$ color indices of the PCSM particles in the initial state are contracted, in a way that the state has no charge. This is a very simple assumption, but one which has far reaching consequences at large $N$. In particular, some form factors of local operators are suppressed by $1/N$ terms, when considering these color neutral states. This greatly simplifies the computation of correlation functions. We conjecture that the standard mass quench may produce this kind of integrable and color neutral initial states.

 In the next section we review the known exact results for the PCSM, including the two-particle S-matrix and some form factors. In Section 3 we determine the structure of integrable and color neutral initial states, keeping only the contributions that lead to nontrivial form factors of operators at infinite $N$. In Section 4 we show a particularly nice property of our initial states, which is that they provide a way to bypass the no-go results of Delfino and Schuricht. These initial states allow for particle transmission from the pre-quench state, and full factorization, while keeping some nontrivial interactions. In Sections 5 and 6 we compute the time-dependent one- and two-point function of the energy-momentum tensor operator after the quench. This computation is done simply by inserting the known form factors of this operator. In Section 7 we study the one- and two-point functions of the renormalized field operator. For this operator, however, we find a trivial result, since all time-dependent contributions are suppressed by $1/N$. We present our conclusions in the last section.

\section{S-matrix and Form factors of the PCSM}

In an integrable theory, like the PCSM, all scattering events are completely elastic and factorizable into a product of two-particle S-matrices. A state containing an elementary excitation is identified by specifying its rapidity $\theta$, related to its energy and momentum by $E=m\cosh\theta$, $p=m\sinh\theta$, its left and right ${\rm SU}(N)$ color indices $a,b=1,\dots,N$, respectively, and distinguishing  whether the excitation is a particle or an antiparticle. The one-particle incoming state and a one-antiparticle incoming states can be written as
\beq
\vert P,\theta,a,b\rangle_{\rm in},\,\,\,\,\,\,\,\,\vert A,\theta,b,a\rangle_{\rm in},\nonumber
\eeq
respectively.

The antiparticle-particle S-matrix, $S(\theta)_{a_1b_1;b_2a_2}^{d_2c_2;c_1d_1}$, defined by 
\beq
\,_{\rm out}\langle A, \theta^\prime_1,d_1,c_1; P,\theta^\prime_2,c_2,d_2\vert A, \theta_1, b_1,a_1; P, \theta_2,a_2,b_2\rangle_{\rm in}=S(\theta)_{a_1b_1;b_2a_2}^{d_2c_2;c_1d_1}\,4\pi \delta(\theta_1-\theta^\prime_1)\,4\pi \delta(\theta_2-\theta^\prime_2),\nonumber
\eeq
 is known to be \cite{wiegmann}
\beq
S(\theta)_{AP}\,_{a_1b_1;b_2a_2}^{d_2c_2;c_1d_1}=Q(\theta)\left[\delta_{a_1}^{c_1}\delta_{a_2}^{c_2}-\frac{2\pi {\rm i}}{N(\pi {\rm i}-\theta)}\delta_{a_1a_2}\delta^{c_1c_2}\right]\left[\delta_{b_1}^{d_1}\delta_{b_2}^{d_2}-\frac{2\pi {\rm i}}{N(\pi {\rm i}-\theta)}\delta_{b_1b_2}b^{d_1d_2}\right],\label{sindex}
\eeq
where
\beq
Q(\theta)=\frac{\sinh\left[\frac{(\pi {\rm i}-\theta)}{2}-\frac{\pi {\rm i}}{N}\right]}{\sinh\left[\frac{(\pi {\rm i}-\theta)}{2}+\frac{\pi {\rm i}}{N}\right]}\,\left\{\frac{\Gamma[i(\pi {\rm i}-\theta)/2\pi+1]\Gamma[-{\rm i}(\pi {\rm i}-\theta)/2\pi-{1}/{N}]}{\Gamma[{\rm i}(\pi {\rm i}-\theta)/2\pi+1-1/N]\Gamma[-{\rm i}(\pi {\rm i}-\theta)/2\pi]}\right\}^2,\label{wiegmann}
\eeq
and $\theta=\theta_1-\theta_2$. The particle-particle and antiparticle-antiparticle S-matrices can be found using crossing symmetry. An incoming particle (antiparticle) can be turned into an outgoing antiparticle (particle), by shifting its rapidity by $\theta\to\theta-\pi i$. 

The planar limit of the PCSM consists of taking $N\to\infty$, while keeping $m$ constant. In this limit, 
the function (\ref{wiegmann}) becomes $Q(\theta)=1+\mathcal{O}\left(1/N^2\right)$.

At infinite $N$, two excitations interact nontrivially only if they have color indices contracted with each other. This is easily seen from Eq. (\ref{sindex}). The terms in the S-matrix, proportional to $\delta_{a_1 a_2}\delta^{c_1 c_2}$, or $\delta_{b_1 b_2}\delta^{d_1 d_2}$, vanish at large $N$, unless one sums over the colors of one of these Kronecker deltas. A particle has a left and a right color index, so it can interact nontrivially with at most two other excitations.

The simplicity of the large-$N$ theory comes from the fact that it can be written as a diagonal scattering theory.
The only nontrivial scattering process is when two excitations have one or two color-index contractions. One can then forget about the color structure and reformulate the problem as a diagonal scattering theory of particles that can interact with the S-matrices
\beq
S_{AP}(\theta)=1, \frac{\theta+\pi i}{\theta-\pi i},\left(\frac{\theta+\pi i}{\theta-\pi i}\right)^2,\label{diagonals}
\eeq
corresponding to whether the two excitations have no color index contractions between them, one contraction, or both colors contracted, respectively.

We will use the form factors of the trace of the energy momentum tensor, $\Theta=T^\mu_\mu$. The form factors of $T_{\mu\nu}$ have been found in Ref. \cite{correlation}. By $\rm{SU}(N)$ color symmetry, non vanishing form factors have the same number of particles and antiparticles. These are, for any positive integer, $M$,
\beq
\langle 0\!\!\!&\vert&\!\!\! T_{\mu\nu}(0)\vert A,\theta_1,b_1,a_1;\dots;A,\theta_M,b_M,a_M;P,\theta_{M+1},a_{M+1},b_{M+1};\dots;P,\theta_{2M},a_{2M},b_{2M}\rangle\nonumber\\
&=&[(p_1+\cdots+p_{2M})_\mu(p_1+\cdots+p_{2M})_\nu-\eta_{\mu\nu}(p_1+\cdots+p_{2M})^2]\nonumber\\
&&\times\frac{1}{N^{M-1}}\sum_{\sigma,\tau\in S_{M}}F^T_{\sigma\tau}(\theta_1,\dots,\theta_{2M})\prod_{j=1}^M\delta_{a_j a_{\sigma(j)+M}}\prod_{k=1}^M\delta_{b_k b_{\tau(k)+M}},\label{tformfactor1}
\eeq
where $\sigma$ and $\tau$ are the permutations that take the numbers $1,\dots,M$ to $\sigma(1),\dots,\tau(M)$ and $\tau(1),\dots,\tau(M)$, respectively. At large $N$, the functions, $F_{\sigma \tau}^T(\{\theta\})$ are
\beq
F^T_{\sigma\tau}(\theta_1,\dots,\theta_M)=\left\{\begin{array}{c}
\frac{(-2\pi^2)(4\pi)^{M-1}}{\prod_{j=1}^M(\theta_j-\theta_{\sigma(j)+M}+\pi i)\prod_{k=1}^M(\theta_k-\theta_{\tau(k)+M}+\pi i)},\,\,{\rm for} \,\sigma(j)\neq\tau(j),\,{\rm for\,all}\,j,\\
0,\,\,{\rm otherwise}\end{array}\right. .\label{tformfactor}
\eeq
Other orderings of the incoming particles can be found by using the S-matrix (\ref{diagonals}).

\section{Integrable and neutral initial states}

\begin{figure}
  \centering
    \begin{tikzpicture}
\draw[line width=3pt] (-6,0) -- (6,0);
\draw (-4.5,2) -- (-4,0);
\draw [->] (-4,0)--(-4.25,1);
\draw (-3.5,2) -- (-4,0);
\draw[<-] (-3.75,1)--(-3.5,2);
\draw (-1.7,2)-- (-1,0);
\draw[->](-1,0)--(-1.35,1);
\draw (-1,0)--(-.3,2);
\draw[<-](-.65,1)--(-1,0);
\draw (2.5,0)--(0.5,2);
\draw[->](0.5,2)--(1.5,1);
\draw (2.5,0)--(4.5,2);
\draw[->](4.5,2)--(3.5,1);
\node at (-6.5,0) { $t=0$};
\node at (-4,-.5) {$K_1(\theta_1)$};
\node at (-1,-.5) {$K_2(\theta_2)$};
\node at (2.5,-.5) {$K_3(\theta_3)$};
\node at (-4.5,2.3) {\small $-\theta_1$};
\node at (-3.5,2.3) {\small $\theta_1$};
\node at (-1.7,2.3) {\small $-\theta_2$};
\node at (-.3,2.3) {\small $\theta_2$};
\node at (0.5,2.3) {\small $-\theta_3$};
\node at (4.5,2.3) {\small $\theta_3$};
\end{tikzpicture}

    \caption{Cooper-pair structure of the initial state. The direction of the arrows denotes if the excitation is a particle (pointing away from the boundary) or an antiparticle (pointing towards the boundary)}
\end{figure}
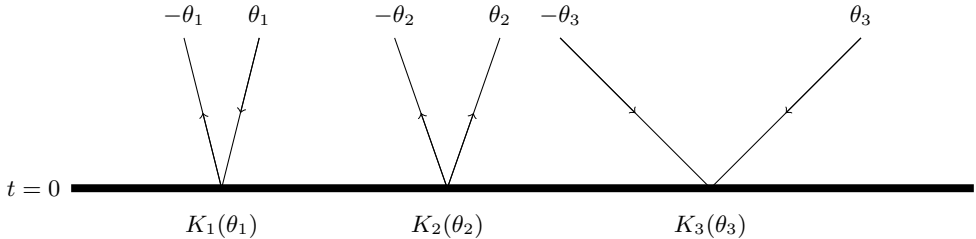

In this section we find what are the initial states of the PCSM that correspond to integrability-preserving boundary conditions, and that are color neutral (that is, all the particle color indices are contracted, forming color singlets). We also use the crucial assumption that the initial state can be described in terms of the massive particles of the low temperature phase, an assumption that is justified for a small quench. In a very highly energetic quench (one where the amount of energy introduced is greater than the critical temperature), one can expect that the initial state involves the massless degrees of freedom of the deconfined phase proposed in \cite{vicari}. 

 Integrable initial states,  are composed of pairs of excitations with  equal mass and opposite rapidities \cite{GZ}, as pictured in Figure 1.  A general integrable neutral state is then given by
\beq
\vert B\rangle&=&\sum_{M=0}^\infty\sum_{\sigma}\vert B_M^{(\sigma)}\rangle=\sum_{M=0}^\infty\sum_{\sigma} \int \left(\prod_{j=1}^M \frac{d\theta_j}{4\pi}\right) \frac{1}{M!}K_\sigma^{C_1C_2\dots C_M D_1 \dots D_M}(\theta_1,\theta_2,\dots,\theta_M)\nonumber\\
&&\times\vert I_1,  -\theta_1,C_1;I_2,-\theta_2,C_2;\dots;I_M,-\theta_M,C_M;J_M,\theta_M,D_M;\dots;J_M,\theta_1,D_1\rangle,\label{boundarystate}
\eeq
where $I_k,J_k=P,A$ label whether the given excitation is a particle or antiparticle, and $C_k$ and $D_k$ denote the set of color indices $a_k,b_k$ for a particle and $b_k,a_k$ for an antiparticle.  The sum over the index $\sigma$ denotes all the possible permutations of color contractions of the excitations, such that the states are color neutral.

There is one more aspect of integrable initial states which we have not imposed, which is that the  state must be factorizable. It must be possible to factorize the function $K(\theta_1,\theta_2,\dots,\theta_M)$ into a product of one-pair functions: $K_1(\theta_1)\times\dots\times K_M(\theta_M)$. These functions $K_k(\theta_k)$ must satisfy the boundary Yang-Baxter equation, which is shown pictorially in Figure 2. There are however some combinations of color contractions in the initial state which would lead to an initial state that doesn't satisfy the boundary Yang-Baxter equation. We will discard these possible color combinations, since they lead to non-integrable initial states.

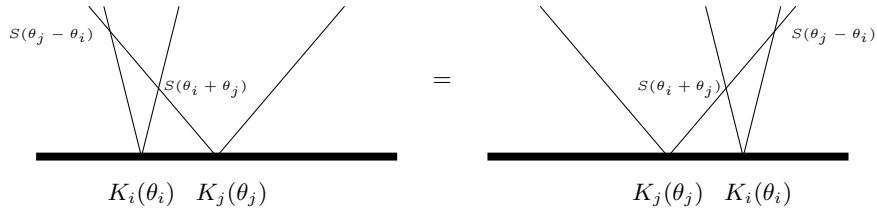
\begin{figure}
  \centering
    \begin{eqnarray}
\begin{tikzpicture}
\draw[line width=3pt] (-2.4,0) -- (2.4,0);
\draw(-1,0)--(-1.5,2);
\draw (-1,0)--(-.5,2);
\draw(0,0)--(-1.7,2);
\draw(0,0)--(1.7,2);
\node at (-1,-.5) {\small $K_i(\theta_i)$};
\node at (0.2,-.5) {\small $K_j(\theta_j)$};
\node at (-2.2,1.6) {\tiny $S(\theta_j-\theta_i)$};
\node at (-.15,.9) {\tiny $S(\theta_i+\theta_j)$};

\node at (3,1) {$=$};

\begin{scope}[shift={(6,0)}];
\draw[line width=3pt] (-2.4,0) -- (2.4,0);
\draw(1,0)--(.5,2);
\draw (1,0)--(1.5,2);
\draw(0,0)--(-1.7,2);
\draw(0,0)--(1.7,2);
\node at (0,-.5) {\small $K_j(\theta_j)$};
\node at (1.2,-.5) {\small $K_i(\theta_i)$};
\node at (2.2,1.6) {\tiny $S(\theta_j-\theta_i)$};
\node at (.15,.9) {\tiny $S(\theta_i+\theta_j)$};

\end{scope}
\end{tikzpicture}
\nonumber
\end{eqnarray}

    \caption{Boundary Yang-Baxter equation}
\end{figure}

One further requirement that we will impose on our initial states, is that we will consider only combinations of colors which lead to nontrivial form factors at large $N$. That is, we sum only over color contractions, $\sigma$, such that the form factors
\beq
\langle0\vert \mathcal{O}\vert B_M^{(\sigma)}\rangle,\nonumber
\eeq
 are not suppressed with subleading powers of $1/N$ for all operators $\mathcal{O}$. This means that the contributions from these form factors to the correlation functions of the operator $\mathcal{O}$ should not be suppressed by a higher order of $1/N$ than the zero temperature correlators. At large $N$, the leading form factors are given by considering states, $\vert B_M^{(\sigma)}\rangle$, where  all the excitations  are connected to each other by color contractions, forming a single string of excitations. These nontrivial particle configurations are shown in Figure 3. 

\begin{figure}
  \centering
    \begin{eqnarray}
\begin{tikzpicture}[scale=.7]
\draw (-2,1)--(-1,2.5);
\draw(-1,2.5)--(1,2.5);
\draw (1,2.5)--(2,1);
\draw (2,1)--(2,-1);
\draw (2,-1)--(1,-2.5);
\draw (1,-2.5)--(-1,-2.5);
\draw(-1,-2.5)--(-2,-1);
\draw (-2,-1)--(-2,1);
\draw (-2,1) circle (5pt);
\fill[black] (-2,1) circle (5pt);
\draw (-2,-1) circle (5pt);
\fill[black!15!white] (-2,-1) circle (5pt);
\draw (-1,2.5) circle (5pt);
\fill[black!15!white] (-1,2.5) circle (5pt);
\draw (1,2.5) circle (5pt);
\fill[black] (1,2.5) circle (5pt);
\draw (2,1) circle (5pt);
\fill[black!15!white] (2,1) circle (5pt);
\draw (2,-1) circle (5pt);
\fill[black] (2,-1) circle (5pt);
\draw (-1,-2.5) circle (5pt);
\fill[black] (-1,-2.5) circle (5pt);
\draw (1,-2.5) circle (5pt);
\fill[black!15!white] (1,-2.5) circle (5pt);
\node at (-3,0) {a)};
\end{tikzpicture}
\,\,\,\,\,\,\,\,\,\,\,\,\,\,\,\,\,\,\,\,\,\,\,\,\,\,\,\,\,\,\,\,\,\,\,\,\,
\begin{tikzpicture}[scale=.7]
\draw (-2,1)--(-1,2.5);
\draw(-1,2.5)--(-1,-2.5);
\draw (-1,-2.5)--(-2,-1);
\draw (-2,-1)--(-2,1);
\draw (2,-1)--(1,-2.5);
\draw (1,-2.5)--(1,2.5);
\draw(1,2.5)--(2,1);
\draw (2,1)--(2,-1);
\draw (-2,1) circle (5pt);
\fill[black] (-2,1) circle (5pt);
\draw (-2,-1) circle (5pt);
\fill[black!15!white] (-2,-1) circle (5pt);
\draw (-1,2.5) circle (5pt);
\fill[black!15!white] (-1,2.5) circle (5pt);
\draw (1,2.5) circle (5pt);
\fill[black] (1,2.5) circle (5pt);
\draw (2,1) circle (5pt);
\fill[black!15!white] (2,1) circle (5pt);
\draw (2,-1) circle (5pt);
\fill[black] (2,-1) circle (5pt);
\draw (-1,-2.5) circle (5pt);
\fill[black] (-1,-2.5) circle (5pt);
\draw (1,-2.5) circle (5pt);
\fill[black!15!white] (1,-2.5) circle (5pt);
\node at (-3,0) {b)};
\end{tikzpicture}
\nonumber
\end{eqnarray}

      \caption{a) The leading contributions to the form factors in the large-$N$ expansion come only from states where all the color indices of the excitations are contracted with other excitations, forming a single closed string. The grey circles represent an antiparticle, while the black circles represent a particle. The pictured contraction of colors gives a leading-order contribution to the eight-excitation form factor. b) This type of color contraction for the eight-excitation form factor, where the particles form more than one closed string, contributes only at higher orders of $1/N$.}
\end{figure}
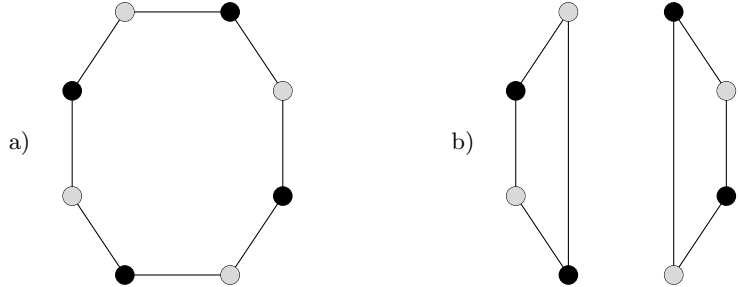

We will keep in the initial state only the combinations of color contractions,  $\sigma$, which satisfy the boundary Yang-Baxter equation, and which produce nontrivial form factors at large $N$. To identify these states,  it is instructive at this point to examine the related theory of the PCSM with a spatial boundary in the $x^1$-direction, rather than a temporal boundary.

It is known that the pair creation amplitude $K(\theta)$ (ignoring for now all color labels) in an initial state is related to the boundary S-matrix, $R(\theta)$, of a theory with a spatial boundary. These functions are related by a shift in the rapidities:
\beq
K(\theta)=R\left(\frac{i\pi}{2}-\theta\right).\nonumber
\eeq
This relation between the temporal and spatial boundaries is explained in Figure 4. We also introduce a new diagrammatic representation for the spatial boundary scattering, which will turn out convenient for classifying the nontrivial initial states.

\begin{figure}
  \centering

\begin{eqnarray}
\begin{tikzpicture}
\begin{scope}[shift={(-4,0)}]
\draw [line width=3pt] (-1.5,-1)--(1.5,-1);
\draw (0,-1)--(-1,1);
\draw (0,-1)--(1,1);
\draw [->](-1,1)--(-.5,0);
\draw [->] (0,-1)--(.5,0);
\draw [dashed] (-.6,.2).. controls (-.5,.4) and (.5,.4) ..(.6,.2);
\draw [dashed] (-.7,.4).. controls (-.6,.6) and (.6,.6) ..(.7,.4);
\node at (-2,-1) {\small $t=0$};
\node at (0,-1.5) {\small $K(\theta)$};
\node at (-2.8,0) {a)};
\end{scope}

\node at (-2,0) {\large $\Rightarrow$};

\begin{scope}[rotate=90]
\draw [line width=3pt] (-1.5,-1)--(1.5,-1);
\draw (0,-1)--(-1,1);
\draw (0,-1)--(1,1);
\draw [->](-1,1)--(-.5,0);
\draw [->] (0,-1)--(.5,0);
\draw [dashed] (-.6,.2).. controls (-.5,.4) and (.5,.4) ..(.6,.2);
\draw [dashed] (-.7,.4).. controls (-.6,.6) and (.6,.6) ..(.7,.4);
\node at (-2,-1) {\small $x=0$};
\node at (0,-1.5) {\small $R(\theta)$};
\end{scope}

\node at (2.8,0) {\large $=$};

\begin{scope} [shift={(5,0)}]
\draw [line width=3pt] (0,-1.5)--(0,1.5);
\draw (-1,0) circle(3.5pt);
\fill[black] (-1,0) circle (3.5pt);
\draw (1,0) circle(3.5pt);
\fill[black] (1,0) circle (3.5pt);
\draw (-1,.1)--(1,.1);
\draw (-1,-.1)--(1,-.1);
\node at (-1.2,-1.3) {\small in};
\node at (1.2,-1.3) {\small out};
\end{scope}

\end{tikzpicture}
\nonumber
\end{eqnarray}

\begin{eqnarray}
\begin{tikzpicture}
\node at (-4,0) {b)};
\draw [line width=3pt] (-3,-1)--(3,-1);
\draw (-1-.3,-1)--(-2-.3,1);
\draw (-1-.3,-1)--(0-.3,1);
\draw [->](0-.3,1)--(-.5-.3,0);
\draw [->] (-1-.3,-1)--(-1.5-.3,0);
\draw [dashed] (-1.6-.3,.2).. controls (-1.5-.3,.4) and (-.5-.3,.4) ..(-.4-.3,.2);

\draw (1+.3,-1)--(0+.3,1);
\draw (1+.3,-1)--(2+.3,1);
\draw [->](1+.3,-1)--(.5+.3,0);
\draw [->] (2+.3,1)--(1.5+.3,0);
\draw [dashed] (.4+.3,.2).. controls (.5+.3,.4) and (1.5+.3,.4) ..(1.6+.3,.2);

\draw [dashed] (-1.7-.3,.4).. controls (-1,2) and (1,2) ..(1.7+.3,.4);
\draw [dashed] (-.3-.3,.4).. controls (-.1,.8) and (.1,.8).. (.3+.3,.4);
\node at (-3.5,-1) {\small $t=0$};
\node at (4,0) {\large $\Rightarrow$};

\begin{scope}[shift={(6.6,0)}]
\draw [line width=3pt] (0,-1.5)--(0,1.5);

\draw(-1,.4)--(1,-.4);
\draw(-1,.6)--(1,.4);
\draw (1,-.6)--(-1,-.4);
\draw (-1,-.6).. controls (-1.5,-.4) and (-1.5,.8).. (-1.5,1);
\draw (-1.5,1).. controls (-1.5, 2) and (1, .7) .. (1,.6);

\draw (-1,.5) circle(3.5pt);
\fill[black! 15! white] (-1,.5) circle (3.5pt);
\draw (1,.5) circle(3.5pt);
\fill[black! 15! white] (1,.5) circle (3.5pt);
\draw (-1,-.5) circle(3.5pt);
\fill[black! 15! white] (-1,-.5) circle (3.5pt);
\draw (1,-.5) circle(3.5pt);
\fill[black! 15! white] (1,-.5) circle (3.5pt);
\node at (-1.2,-1.3) {\small in};
\node at (1.2,-1.3) {\small out};
\end{scope}

\end{tikzpicture}
\nonumber
\end{eqnarray}

      \caption{a) An antiparticle-particle  pair creation amplitude is related to the boundary S-matrix of particle interacting with a spatial boundary and exiting as a particle. Each dashed line denotes that one color index is contracted between the two excitations it connects (in this case both colors are contracted). In the right-hand side of a) we introduce a new kind of diagram to represent the color index contractions of excitations when they scatter with a spatial boundary, where on the ``in" side we draw the different excitations that are approaching the boundary before scattering with it, and on the ``out" side we draw all the excitations after scattering with the boundary. The lines in these new diagrams represent the color contractions between the excitations. b) Left-hand side: one typical initial state with two  pairs and some combination of color contractions; right-hand side: the representation of this state in our new diagrammatic method.}
\end{figure}
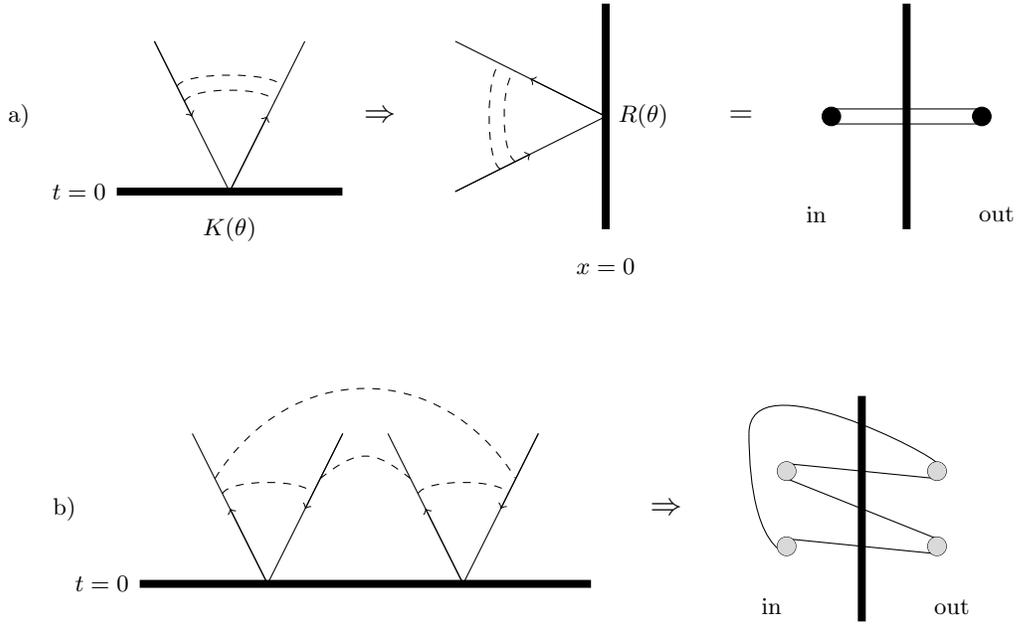

In the theory with a spatial boundary, an excitation interacts nontrivially with the boundary, only if it has one or both of its color indices contracted with the boundary.  If an excitation instead has its two color indices contracted with other particles, and not the boundary, the scattering with the boundary is the same as that of a free boson.

We are now ready to identify which are the factorizable nontrivial initial states. These states correspond in the spatial boundary picture, to the scattering of an open string of excitations with the boundary. That is, there is a string of alternating particles and antiparticles, each excitation has one of its indices contracted with its nearest neighbor. The two excitations at the end of the string have one free color index each, these color indices are contracted with the boundary.  This type of string scattering is pictured in Figure 5. Given a string of $M$ excitations scattering with a boundary, there are $2(M!)$ possible color contractions which lead to nontrivial and factorizable scattering states.

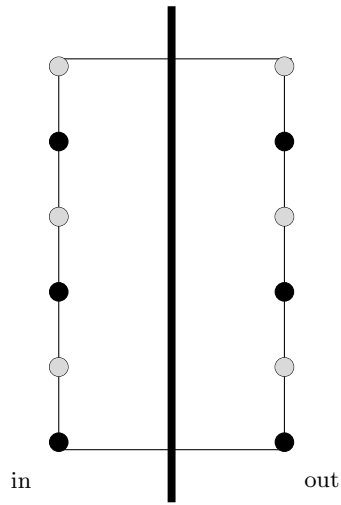
\begin{figure}
  \centering

\begin{eqnarray}
\begin{tikzpicture}

\draw (-1.5,2.5)--(-1.5,1.5);
\draw (-1.5,1.5)--(-1.5,.5);
\draw (-1.5,.5)--(-1.5,-.5);
\draw (-1.5,-.5)--(-1.5,-1.5);
\draw (-1.5,-1.5)--(-1.5,-2.5);
\draw (1.5,2.5)--(1.5,1.5);
\draw (1.5,1.5)--(1.5,.5);
\draw (1.5,.5)--(1.5,-.5);
\draw (1.5,-.5)--(1.5,-1.5);
\draw (1.5,-1.5)--(1.5,-2.5);
\draw(-1.5,2.6)--(1.6,2.6);
\draw(-1.5,-2.6)--(1.5,-2.6);

\draw [line width=3pt] (0,-3.3)--(0,3.3);
\draw (-1.5,2.5) circle(3.5pt);
\fill[black! 15! white] (-1.5,2.5) circle (3.5pt);
\draw (-1.5,1.5) circle(3.5pt);
\fill[black] (-1.5,1.5) circle (3.5pt);
\draw (-1.5,.5) circle(3.5pt);
\fill[black! 15! white] (-1.5,.5) circle (3.5pt);
\draw (-1.5,-.5) circle(3.5pt);
\fill[black] (-1.5,-.5) circle (3.5pt);
\draw (-1.5,-1.5) circle(3.5pt);
\fill[black! 15! white] (-1.5,-1.5) circle (3.5pt);
\draw (-1.5,-2.5) circle(3.5pt);
\fill[black] (-1.5,-2.5) circle (3.5pt);

\draw (1.5,2.5) circle(3.5pt);
\fill[black! 15! white] (1.5,2.5) circle (3.5pt);
\draw (1.5,1.5) circle(3.5pt);
\fill[black] (1.5,1.5) circle (3.5pt);
\draw (1.5,.5) circle(3.5pt);
\fill[black! 15! white] (1.5,0.5) circle (3.5pt);
\draw (1.5,-.5) circle(3.5pt);
\fill[black] (1.5,-.5) circle (3.5pt);
\draw (1.5,-1.5) circle(3.5pt);
\fill[black! 15! white] (1.5,-1.5) circle (3.5pt);
\draw (1.5,-2.5) circle(3.5pt);
\fill[black] (1.5,-2.5) circle (3.5pt);

\node at (-2,-3) {\small in};
\node at (2,-3) {\small out};
\end{tikzpicture}
\nonumber
\end{eqnarray}

      \caption{Representation of a six Cooper-pair initial state that lead to nontrivial form factors, and which satisfies the boundary Yang-Baxter equation. This type of contributions can always be drawn as a simple square, with alternating particles and antiparticles connected to their nearest neighbors, with only two color lines crossing the boundary from the ``in" to the ``out" side. There are $2(M!)$ diagrams of this type which contain $M$ Cooper pairs.}
\end{figure}

We now want to verify that the initial state that corresponds to the string scattering we have just described, in fact satisfies the boundary Yang-Baxter equation. As an example, we show in Figure 6 the initial state corresponding to a 4-excitation string scattering. This state is given by
\beq
&&\int \left(\prod_{j=1}^4\frac{d\theta_j}{4\pi} \right)\frac{1}{4!}K_1(\theta_1) K_2(\theta_2)K_3(\theta_3)K_4(\theta_4)\,\delta_{a_1a_2}\delta_{b_2b_3}\delta_{a_3a_4}\delta_{b_4d_4}\delta_{c_4c_3}\delta_{d_3d_2}\delta_{c_2c_1}\delta_{d_1b_1}\nonumber\\
&&\times\vert P,-\theta_1,a_1,b_1;A,-\theta_2,b_2,a_2;P,-\theta_3,a_3,b_3;A,-\theta_4,b_4,a_4;\nonumber\\
&&\,\,\,\,\,\,\,\,\,\,\,\,P,\theta_4,c_4,d_4;A,\theta_3,d_3,a_3;P,\theta_2,c_2,d_2;A,\theta_1,d_1,c_1\rangle.
\eeq

\begin{figure}
  \centering
   
\begin{eqnarray}
\begin{tikzpicture}

\draw (-1.5,1.5)--(-1.5,.5);
\draw (-1.5,.5)--(-1.5,-.5);
\draw (-1.5,-.5)--(-1.5,-1.5);
\draw (1.5,1.5)--(1.5,.5);
\draw (1.5,.5)--(1.5,-.5);
\draw (1.5,-.5)--(1.5,-1.5);
\draw (-1.5,1.6)--(1.5,1.6);
\draw (-1.5,-1.6)--(1.5,-1.6);

\draw [line width=3pt] (0,-2.3)--(0,2.3);
\draw (-1.5,1.5) circle(3.5pt);
\fill[black] (-1.5,1.5) circle (3.5pt);
\draw (-1.5,.5) circle(3.5pt);
\fill[black! 15! white] (-1.5,.5) circle (3.5pt);
\draw (-1.5,-.5) circle(3.5pt);
\fill[black] (-1.5,-.5) circle (3.5pt);
\draw (-1.5,-1.5) circle(3.5pt);
\fill[black! 15! white] (-1.5,-1.5) circle (3.5pt);

\draw (1.5,1.5) circle(3.5pt);
\fill[black] (1.5,1.5) circle (3.5pt);
\draw (1.5,.5) circle(3.5pt);
\fill[black! 15! white] (1.5,.5) circle (3.5pt);
\draw (1.5,-.5) circle(3.5pt);
\fill[black] (1.5,-.5) circle (3.5pt);
\draw (1.5,-1.5) circle(3.5pt);
\fill[black! 15! white] (1.5,-1.5) circle (3.5pt);

\node at (-2,-2) {\small in};
\node at (2,-2) {\small out};
\node at (-2.5,0) { a)};

\end{tikzpicture}
\nonumber
\end{eqnarray}

\begin{eqnarray}
\begin{tikzpicture}[scale=.9]
\draw [line width=3pt] (-7.3,-1)--(7.3,-1);
\draw (-6,-1)--(-7,1);
\draw [->] (-6,-1)--(-6.5,0);
\draw (-6,-1)--(-5,1);
\draw[->] (-5,1)--(-5.5,0);
\draw(-2,-1)--(-3,1);
\draw[->] (-3,1)--(-2.5,0);
\draw (-2,-1)--(-1,1);
\draw[->] (-2,-1)--(-1.5,0);
\draw(2,-1)--(1,1);
\draw[->] (2,-1)--(1.5,0);
\draw(2,-1)--(3,1);
\draw[->] (3,1)--(2.5,0);
\draw (6,-1)--(5,1);
\draw[->] (5,1)--(5.5,0);
\draw(6,-1)--(7,1);
\draw[->](6,-1)--(6.5,0);

\draw[dashed] (-6.6,.2).. controls (-6.5,.3) and (-5.5,.3)..(-5.4,.2);
\draw[dashed] (6.6,.2).. controls (6.5,.3) and (5.5,.3)..(5.4,.2);
\draw[dashed] (-5.3,.4).. controls (-4,2) and (-2,1.5)..(-1.6,-.2);
\draw[dashed] (-6.7,.4).. controls (-6,2) and (-4,2)..(-2.4,-.2);
\draw[dashed] (-2.7,.4).. controls (-2,2) and (0,2).. (1.6,-.2);
\draw[dashed] (-1.3,.4).. controls (0,2) and (2,2).. (2.4,-.2);
\draw[dashed] (1.3,.4).. controls (2,2) and (4,2).. (5.6,-.2);
\draw[dashed] (2.7,.4).. controls (4,2) and (6,2)..(6.7,.4);

\node at (-7.8,-1) {\small $t=0$};
\node at (-8.3,0) { b)};
\node at (-6,-1.5) {\small $K_1(\theta_1)$};
\node at (-2,-1.5) {\small $K_2(\theta_2)$};
\node at (2,-1.5) {\small $K_3(\theta_3)$};
\node at (6,-1.5) {\small $K_4(\theta_4)$};

\end{tikzpicture}
\nonumber
\end{eqnarray}

      \caption{a) A nontrivial four-excitation boundary scattering. b) The corresponding initial state for the same color contractions pictured in a).}
\end{figure}
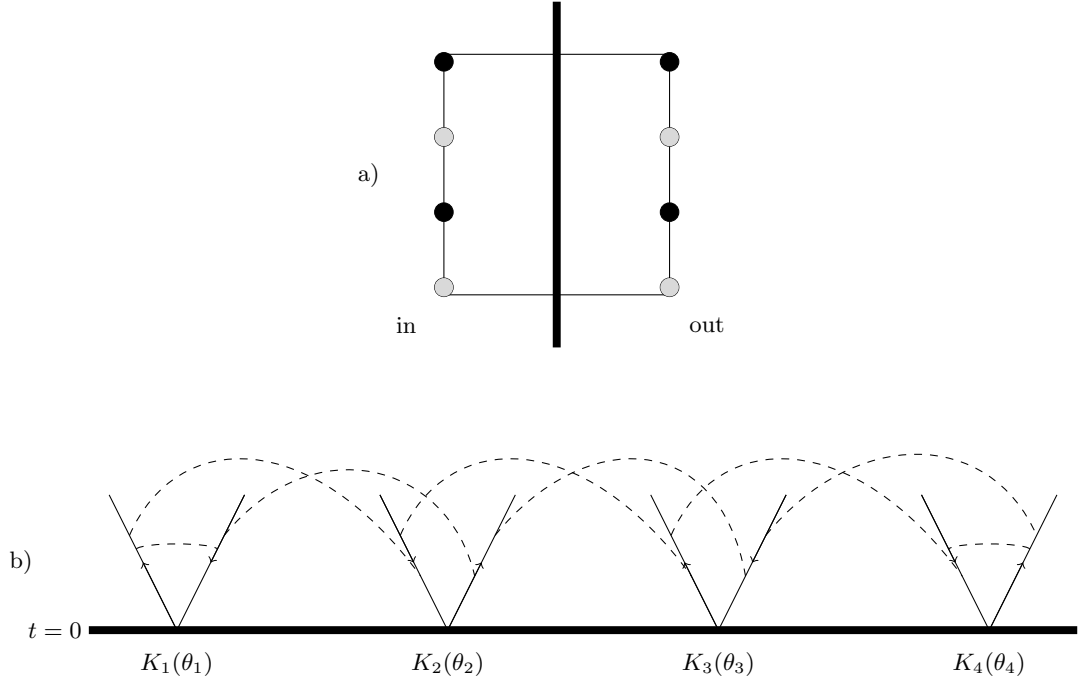

Satisfying the Yang-Baxter equation means that one can permute the order of the creation of the Cooper pairs and the state will be the same. For example, the boundary Yang-Baxter equation for exchanging the pairs 1 and 2 in Figure 6, gives
\beq
S_{PA}(-\theta_1+\theta_2)K_1(\theta_1)K_2(\theta_2)=K_1(\theta_1)K_2(\theta_2)S_{PA}(\theta_2-\theta_1),
\eeq
which is trivially satisfied. This exchange of the pairs 1 and 2 is pictured in Figure 7. It is then very simple to show this way that the boundary Yang-Baxter equation is satisfied for the exchange of any two of the pairs in the initial state.

\begin{figure}
  \centering

\begin{eqnarray}
\begin{tikzpicture}

\draw[line width=3pt] (-2.4,0) -- (2.4,0);
\draw(-1,0)--(-1.5,2);
\draw (-1,0)--(-.5,2);
\draw(0,0)--(-1.7,2);
\draw(0,0)--(1.7,2);

\fill[black] (-1.425,1.67) circle (2pt);
\node at (-1,-.5) {\tiny $K_1(\theta_1)$};
\node at (0,-.5) {\tiny $K_2(\theta_2)$};

\node at (3.25,1) {$=$};

\begin{scope}[shift={(6.5,0)}]
\draw[line width=3pt] (-2.4,0) -- (2.4,0);
\draw(1,0)--(.5,2);
\draw (1,0)--(1.5,2);
\draw(0,0)--(-1.7,2);
\draw(0,0)--(1.7,2);
\fill[black] (1.425,1.67) circle (2pt);
\node at (0,-.5) {\tiny $K_2(\theta_2)$};
\node at (1,-.5) {\tiny $K_1(\theta_1)$};
\end{scope}

\end{tikzpicture}
\nonumber
\end{eqnarray}

      \caption{Boundary Yang-Baxter equation for the the pairs 1 and 2 of Figure 6.b. In each of the two sides of the equation, there is only one nontrivial S-matrix, denoted by the black circle, the other S-matrix usually present in the boundary Yang-Baxter equation is simply 1 in this case.}
\end{figure}
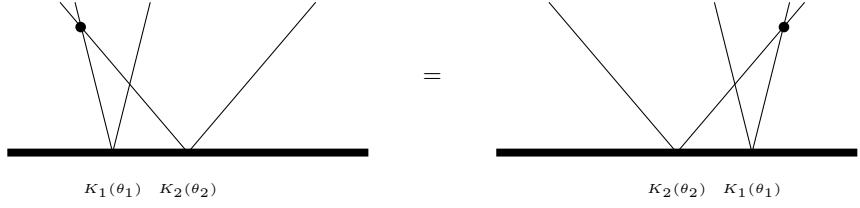

Now that we have found the structure of the initial states that lead to nontrivial form factors, and that satisfy the boundary Yang-Baxter equation, we  want to fix the functions $K_j(\theta_j)$. There is an additional constraint one can impose on integrable boundary conditions, which restricts these functions. This is the cross unitarity condition, pictured in figure 8.

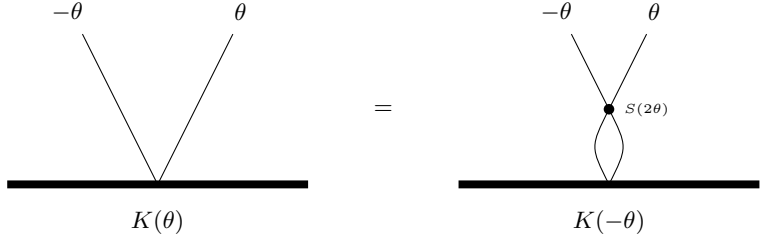
\begin{figure}
  \centering

\begin{eqnarray}
\begin{tikzpicture}
\draw[line width=3pt] (-2,0) -- (2,0);
\draw (0,0)--(-1,2);
\draw(0,0)--(1,2);

\node at (-1.2,2.3) {\small $-\theta$};
\node at (1.1,2.3) {\small $\theta$};
\node at (0,-.5) {\small $K(\theta)$};

\node at (3,1) {$=$};

\begin{scope}[shift={(6,0)}]
\draw[line width=3pt] (-2,0) -- (2,0);
\node at (0,-.5) {\small $K(-\theta)$};

\draw (0,1)--(-.5,2);
\draw(0,1)--(.5,2);
\draw(0,0).. controls (-.25,.5) and (-.25,.5).. (0,1);
\draw(0,0).. controls (.25,.5) and (.25,.5).. (0,1);
\fill[black] (0,1) circle (2pt);
\node at (.5,1) {\tiny $S(2\theta)$};
\node at (-.7,2.3) {\small $-\theta$}; 
\node at (.6,2.3) {\small $\theta$}; 
\end{scope}
\end{tikzpicture}
\nonumber
\end{eqnarray}

      \caption{The cross-unitarity relation.}
\end{figure}

 Again, as an example we consider the four-pair contribution to the initial state, from figure 6. First, it is easy to find the functions $K_2(\theta_2)$ and $K_3(\theta_3)$, of the pairs in the middle of the string.  In this case, the two excitations in the pair 2 (the ones with rapidities $-\theta_2$ and $\theta_2$) do not interact with each other, as they do not have any color indices contracted with each other. Therefore, the cross unitarity condition for this pair, simply gives
 \beq
 K_2(\theta_2)=K_2(-\theta).\label{freecrossunitarity}
 \eeq
The same reasoning applies for the pair 3. Eq. (\ref{freecrossunitarity}) is the cross unitarity relation  satisfied by a theory of free bosons. This means we can write
\beq
K_2(\theta)=K_3(\theta)=\frac{K_{\rm boson}(\theta)}{N},
\eeq
where $K_{\rm boson}(\theta)$ is the pair creation amplitude for a free boson. We have also included a normalization with $1/N$, which is necessary so that we can have a normalizable initial state ($\langle B_M^{(\sigma)}\vert B_M^{(\sigma)}\rangle$ doesn't diverge with $N$). The function $K_{\rm boson}(\theta)$ is known exactly for a quantum quench of a free bosonic theory, where the mass parameter is suddenly changed from $m_0$ to $m$, where
\beq
K_{\rm boson}(\theta)=\frac{\sinh(\theta-\xi)}{\sinh(\theta+\xi)},\label{freeboson}
\eeq
and 
\beq
m_0\sinh \xi=m \sinh \theta,\nonumber
\eeq
is found by a simple Bogoliubov transformation \cite{FM2}.
We conjecture that  this solution for $K_{\rm boson}(\theta)$ may be appropriate to describe the mass quench of the PCSM (though this is not a statement we have proven at this point).

Now we consider the consequences of cross unitarity for the pairs of excitations at the end of the string, namely $K_1(\theta_1)$ and $K_4(\theta_4)$ in figure 6. In this case the two particles in pair 1 (and pair 4) do interact with each other, cross unitarity for the pair 1 gives the relation
\beq
K_1(\theta)=\left(\frac{2\theta-\pi i}{2\theta+\pi i}\right) K_1(-\theta).\label{crosunitk1}
\eeq
The solution to Eq. (\ref{crosunitk1}) is
\beq
K_1(\theta)=\frac{K_{\rm fermion}(\theta)}{N(2\theta+\pi i)},\nonumber
\eeq
where $K_{\rm fermion}(\theta)$ is the pair creation amplitude for free fermions, which satisfies
\beq
K_{\rm fermion}(\theta)=-K_{\rm fermion}(-\theta).
\eeq
The function $K_{\rm fermion}(\theta)$ is also known for a quantum quench corresponding to a change in the mass parameter, and it is given by
\beq
K_{\rm fermion}(\theta)=i\frac{\sinh\left(\frac{\theta-\xi}{2}\right)}{\sinh\left(\frac{\theta+\xi}{2}\right)}.\label{freefermion}
\eeq
With the same method we can easily find 
\beq
K_4(\theta)=\frac{K_{\rm fermion}(\theta)}{N(2\theta-\pi i)}.
\eeq

Now it is easy to generalize our results to any generic contribution to the initial state, corresponding to an $M$ excitation string, for all $M>1$. All the pairs of excitations in the bulk of the string will be created with the amplitude $K_j(\theta)=K_{\rm boson}(\theta)/N$, for $1<j<M$. The pairs at the ends of the string, namely the pairs 1 and $M$, will be created with the amplitude
\beq
K_1,K_M(\theta)=\frac{K_{\rm fermion}(\theta)}{N(2\theta\pm \pi i)},\nonumber
\eeq
where the $\pm$ depends on whether the excitation on the left side of the pair (the one with rapidity $-\theta$) is a particle or an antiparticle. This completes the determination of all the integrable nontrivial initial states for $M>1$.  The $M=1$ contribution can also be easily computed, in this case the one pair creation amplitude satisfies the cross unitarity condition
\beq
K_{M=1}(\theta)=\left(\frac{2\theta\mp\pi i}{2\theta\pm\pi i}\right)^2K_{M=1}(-\theta),\nonumber
\eeq
with solution
\beq
K_{M=1}(\theta)=\frac{K_{\rm boson}(\theta)}{N(2\theta\pm i\pi)^2}.\nonumber
\eeq

We conjecture that the initial states we have found corresponds to, or at least approximates the initial state after a standard mass quench of the PCSM. A sudden change in the particle mass should produce a color neutral initial state, if the pre-quench state was also color neutral. In particular, we are able to write the pair creation amplitudes in terms of the free boson and free fermion amplitudes, which can be found for the genuine mass quench by a Bogoliubov transformation. As we will see in the next section, our initial state also allows for particle transmission from the pre-quench state, which is a desirable property in a genuine integrable mass quench.

The appearance of both, $K_{\rm boson}(\theta)$ and $K_{\rm fermion}(\theta)$ can be motivated by by examining the S-matrices at large $N$, Eq. (\ref{diagonals}). The first and third S-matrices in (\ref{diagonals}) correspond to particles with bosonic statistics, which can be seen from the fact that $S(0)=1$, so two particles with the same rapidity commute. The second S-matrix in (\ref{diagonals}) corresponds to particles which obey Fermi-Dirac statistics, seen from the fact that $S(0)=-1$, which means that two particles with the same rapidity anti commute. Naturally, when we apply the cross-unitarity condition, the pair creation amplitudes corresponding to these S-matrices can be easily expressed in terms of $K_{\rm boson}(\theta)$ for the first and third S-matrices in (\ref{diagonals}), and $K_{\rm fermion}(\theta)$ for the second S-matrix. 

Our initial state, expressed in terms of the general functions $K_{\rm boson}(\theta)$ and $K_{\rm fermion}(\theta)$,  is the most general nontrivial, integrable and color-neutral state.  There is however no guarantee that it corresponds to a genuine mass quench in the PCSM. Our choice of initial state might also be motivated by a calculation similar to the one done for the Sinh-Gordon model in Ref.\cite{Spyros}. For this, one would need to use the renormalized field form factors from \cite{renormalizedfield} to express the pre-quench particle- and antiparticle-creation and annihilation operators in terms of the field operator. One can then identify the initial state as that which is annihilated by the pre-quench annihilation operators, and solve this condition in the post-quench particle basis.  We adopt the point of view of simply imposing this initial state, and following the unitary time evolution. The question of whether this kind of initial state corresponds to a mass quench is beyond the scope of this paper.

\section{Factorization and particle transmission}

In this section we discuss the possibility of transmission of pre-quench particles into the post-quench state. It has been shown that in theories with one species of particle, it is impossible to set up a quantum quench which is integrable at all times, which allows particle transmission, and where the post-quench particles interact nontrivially \cite{Delfino}. This statement was later generalized to other non diagonal scattering of theories with many species of particles \cite{Schuricht}.

The proof of these no-go  theorems is based on the consequences of demanding factorization at all times. We will call $T(\theta)$  the amplitude for the transmission of a particle from the pre-quench state, which has rapidity $\theta$ in the post-quench state (we impose that the momentum of the particle is conserved at $t=0$, thus the rapidity changes because there is a change in particle mass, we are ignoring the color indices of the particle for simplicity). This amplitude is represented pictorially in figure 9. 

\begin{figure}
  \centering

\begin{eqnarray}
\begin{tikzpicture}
\draw[line width=3pt] (-2,0) -- (2,0);
\draw (-.7,-1.4)--(0,0);
\draw (.7,1.4)--(0,0);
\draw[->] (-.7,-1.4)--(-.5,-1);
\draw[->] (0,0)--(.5,1);
\fill[black! 70! white] (0,0) circle (4pt);

\node at (-2.5,0) {\small $t=0$};
\node at (-.7,-1.7) {\small $\xi$};
\node at (.8,1.7) {\small $\theta$};
\node at (.35,-.35) {\small $T(\theta)$};

\end{tikzpicture}
\nonumber
\end{eqnarray}

      \caption{Particle transmission amplitude from the pre-quench state to the post-quench state. We assume the momentum of the particle is the same before and after the quench, but since there is a change in the particle mass, the rapidity is different, given by $\xi$ before the quench and $\theta$ afterwards.}
\end{figure}
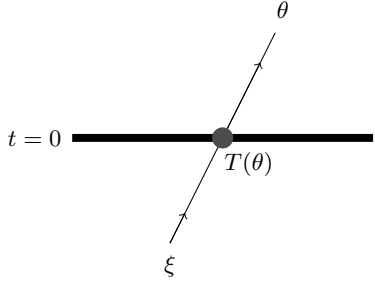

Requiring factorizability at all times, together with the fact that we demanded integrable initial states composed of Cooper pairs, gives us two new consistency relations that must be satisfied by an integrable quench. Suppose we have a particle of rapidity $\theta$, being transmitted from the pre-quench state, interacting with a Cooper pair with amplitude given by $K_i(\theta_i)$. If $0<\theta<\theta_i$, factorizability implies the relation pictured in figure 10.a, which gives us the condition
\beq
S_T(\theta+\theta_i)T(\theta)K(\theta_i)=S_T(\theta_i-\theta)K(\theta_i)T(\theta),\label{condition1}
\eeq
where we have again, suppressed the color indices for simplicity, and $S_T$ is the S-matrix of the transmitted particle with the particles of the Cooper pair. If $0<\theta_i<\theta$, as pictured in Figure 10.b, we get the consistency condition
\beq
S_T(\theta-\theta_i)S_T(\theta+\theta_i)T(\theta)K(\theta_i)=K(\theta_i)T(\theta).\label{condition2}
\eeq

\begin{figure}
  \centering
    

\begin{eqnarray}
\begin{tikzpicture}
\node at (-2.5,1) {a)};

\draw[line width=3pt] (-2,0) -- (2,0);
\draw(0,0)--(-1,2);
\draw(0,0)--(1,2);
\draw(-1,-.5)--(-.6,2);

\fill[black! 70! white] (-.915,0) circle (3pt);
\fill[black] (-.695,1.4) circle (2pt);

\node at (-.915+.3,-.3) {\small $T$};
\node at (-.695-.3,1.4) {\small $S_T$};
\node at (0,-.3) {\small $K$};

\node at (2.75,1) {$=$};

\begin{scope}[shift={(5.5,0)}]
\draw[line width=3pt] (-2,0) -- (2,0);
\draw(0,0)--(-1,2);
\draw(0,0)--(1,2);
\draw(-1+1.3,-.5)--(-.6+1.3,2);

\fill[black! 70! white] (-.915+1.3,0) circle (3pt);
\fill[black] (-.73+1.3,1.12) circle (2pt);

\node at (-.915+1.3+.3,-.3) {\small $T$};
\node at (-.73-.3+1.3,1.12) {\small $S_T$};
\node at (-.1,-.3) {\small $K$};
\end{scope}

\end{tikzpicture}
\nonumber
\end{eqnarray}

\begin{eqnarray}
\begin{tikzpicture}

\node at (-2.5,1) {b)};
\draw[line width=3pt] (-2,0) -- (2,0);
\draw(0,0)--(-1,2);
\draw(0,0)--(1,2);
\draw(-1,-.5)--(1.4,2);

\fill[black! 70! white] (-.5,0) circle (3pt);
\fill[black] (-.185,.35) circle (2pt);
\fill[black] (.57,1.125) circle (2pt);

\node at (-.5,-.3) {\small $T$};
\node at (-.3-.2,.35) {\small $S_T$};
\node at (0,-.3) {\small $K$};
\node at (.57+.3,1.125) {\small $S_T$};

\node at (2.5,1) {$=$};

\begin{scope}[shift={(5,0)}]
\draw[line width=3pt] (-2,0) -- (3,0);
\draw(0,0)--(-1,2);
\draw(0,0)--(1,2);
\draw(-1+1.5,-.5)--(1.4+1.5,2);

\fill[black! 70! white] (-.5+1.5,0) circle (3pt);

\node at (-.5+1.5,-.3) {\small $T$};
\node at (0,-.3) {\small $K$};
\end{scope}

\end{tikzpicture}
\nonumber
\end{eqnarray}

      \caption{Two relations resulting from demanding factorizability at all times.}
\end{figure}
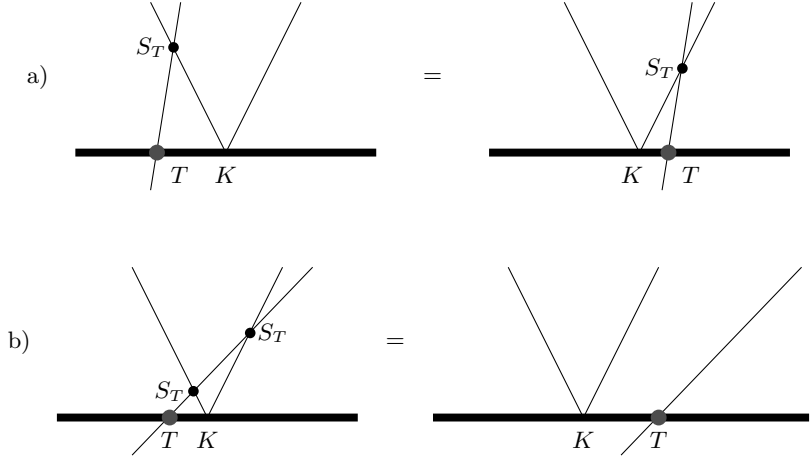

The conclusion from Refs.\cite{Delfino,Schuricht} is that the only way to satisfy both conditions (\ref{condition1}) and (\ref{condition2}), is if $T(\theta)=0$, or $S_T(\theta)=\pm 1$. That is, either there is no particle transmission from the pre-quench state, or one has a free scattering theory. 

In our particular large-$N$ quench, with the initial states discussed in last section, we always have $S_T(\theta)=1$. This is because we demanded that the initial state is colorless, so the particles in the Cooper pair cannot interact nontrivially with the transmitted particle, because they do not have any remaining color indices to contract with it. The conditions (\ref{condition1}) and (\ref{condition2}) are then trivially satisfied.

The subtle distinction to make about our large-$N$ case is that, even though $S_T(\theta)=1$, this doesn't mean the post-quench theory is a free scattering theory. The post-quench particles can still interact with each other, and the form factors of operators are not trivial. In this way, the large-$N$ limit is a way to work around the no-go theorems proved in \cite{Delfino,Schuricht}, since we are able to set up a quantum quench that has both, particle transmission from the pre-quench state, and nontrivial interactions among the post-quench particles.

\section{One-point function of the energy-momentum tensor}

In this section we will compute the one-point function of the trace of the energy-momentum tensor  after the quench, which is given by 
\beq
\frac{1}{N}\frac{\langle B\vert \Theta (t)\vert B\rangle}{\langle B\vert B\rangle}.\label{onepoint}
\eeq
We have chosen to evaluate this particular operator because all its form factors are known, and they are nonzero for color neutral states, such as $\vert B\rangle$. Other operators whose form factors are known are the renormalized field operator, and the $SU(N)$ Noether currents, however, these form factors are zero for color neutral states.

To evaluate the expression (\ref{onepoint}), we need the  form factors with excitations in the incoming and outgoing state. These can be found from the form factors with excitations only in the incoming state using the generalized crossing formula. Using  notation similar to that from Refs.\cite{smirnovbook, leclairmussardo}, suppose $\mathcal{A}=\{I_n,\theta_n,C_n;I_{n-1},\theta_{n-1},C_{n-1};\dots;I_1,\theta_1,C_1\}$, with $\{\theta\}$ forming a set of ordered rapidities, the indices $\{I\}$ denote if the excitation is a particle or an antiparticle, and $\{C\}$ denote the left and right color indices of each excitation. We similarly define $\mathcal{B}=\{I^\prime_1,\theta^\prime_1,C^\prime_1;\dots;I^\prime,\theta^\prime_m,C^\prime_m\}.$ We can then write the general form factor as
\beq
\langle \mathcal{A}\vert \Theta\vert \mathcal{B}\rangle=\sum_{\mathcal{A}=\mathcal{A}_1\cup\mathcal{A}_2;\mathcal{B}=\mathcal{B}_1\cup\mathcal{B}_2}S_{\mathcal{A},\mathcal{A}_1}S_{\mathcal{B},\mathcal{B}_1}\langle\mathcal{A}_2\vert\mathcal{B}_2\rangle\langle\mathcal{A}_1^+\vert\Theta\vert\mathcal{B}_1\rangle,\label{connected}
\eeq
where $S_{\mathcal{A},\mathcal{A}_1}$ is the product of S-matrices required to bring the state $\vert \mathcal{A}\rangle$ into the order $\vert \mathcal{A}_2\mathcal{A}_1\rangle$, and similarly for $S_{\mathcal{B},\mathcal{B}_1}$ and $\vert \mathcal{B}\rangle$.
The products $\langle\mathcal{A}_2\vert\mathcal{B}_2\rangle$ can be evaluated by introducing particle-creation operators, as was shown in \cite{leclairmussardo}. These products are generally infrared divergent, and can be regularized by placing the theory on a finite volume $L$. These divergent products, however, are canceled by similar terms in the denominator, $\langle B\vert B\rangle$. The on the right hand side of (\ref{connected}) are  defined by (suppressing the $\{I\}$ and $\{C\}$ indices for simplicity)
\beq
\langle \theta_n^+;\dots; \theta_1^+\vert \Theta\vert \theta^\prime_1;\dots;\theta^\prime_m\rangle= \langle 0\vert \Theta\vert \theta^\prime_1;\dots;\theta^\prime_m;\theta_n-\pi i+i\eta_n;\dots;\theta_1-\pi i+i\eta_1\rangle,\nonumber
\eeq
with some small numbers, $\eta_j$, that we eventually want to take to zero. This limit of $\eta_j\to0$ has to be taken carefully, as there are annihilation poles that will make the form factors divergent for some values of the rapidities, and a useful regularization approach was proposed in \cite{leclairmussardo}. We will, however, not run into these difficulties, since as we will see, any contributions to the corralation functions from form factors where one would have to integrate over annihilation poles, is suppressed by higher powers of $1/N$ in our problem. 

There is one great simplification that allows us to compute the function (\ref{onepoint}) exactly in the large-$N$ limit. The leading contribution in orders of $1/N$ to the connected form factors of the operator $\Theta$ are those corresponding to states where all the particles are connected to each other with index contractions, forming a single closed string.
 This means that form factors like 
\beq
\frac{1}{N}\langle0\vert \Theta(t)\vert B_M^{(\sigma)}\rangle,\nonumber
\eeq
contribute at leading order of $1/N$, because we have chosen the states $\vert B_M^{(\sigma)}\rangle$ such that the $2M$ excitations form a single closed string. Form factors of the form
\beq
\frac{1}{N}\langle B_K^{(\sigma^\prime)}\,^+\vert \Theta(t)\vert B_M^{(\sigma)}\rangle,\nonumber
\eeq
contribute only at higher orders of the $1/N$ expansion, because the particles in $\langle B_K^{(\sigma^\prime)}\vert$ and $\vert B_M^{(\sigma)}\rangle$ make two separate closed strings.
The leading $1/N$-terms of the function (\ref{onepoint}) can then be written as
\beq
\lim_{N\to\infty}\frac{1}{N}\frac{\langle B\vert \Theta(t)\vert B\rangle}{\langle B\vert B\rangle}&=&\frac{1}{N}\langle B\vert \Theta(t)\vert0\rangle+\frac{1}{N}\langle0\vert \Theta(t)\vert B\rangle\nonumber\\
&=&\sum_{M=0}^\infty\sum_{\sigma} \frac{1}{N}\left[\langle 0\vert \Theta (t)\vert B_M^{(\sigma)}\rangle+\left(\langle 0\vert \Theta (t)\vert B_M^{(\sigma)}\rangle\right)^*\right],
\eeq
where we have used the fact that the factors arising from $\langle \mathcal{A}_2\vert\mathcal{B}_2\rangle$ in (\ref{connected}) cancel exactly with similar factors in the denominator $\langle B\vert B\rangle$.

We now can evaluate each of the contributions, $\langle 0\vert \Theta(t)\vert B_M^{(\sigma)}\rangle$, using the exact form factors of $\Theta$. At this point we recall that for the $M$-Cooper pair contribution, there are $M!$ different contributions denoted by the permutations, $\sigma$, corresponding to the $2(M!)$ possible ways we can make a string of the type described in Figure 5. Each one of these $2(M!)$ configurations actually give the same contribution  to $\langle 0\vert \Theta(t)\vert B\rangle$, once one integrates over all the rapidities, so it suffices to look at only one of these strings, and then multiply the contribution by $2M!$. This factor of $M!$ cancels with the similar factor that was already present in the expression (\ref{boundarystate}). Now, simply using the form factors (\ref{tformfactor1}) we can write
\beq
&\sum_\sigma\frac{1}{N}\langle0\vert \Theta(t)\vert B_M^{(\sigma)}\rangle=4\pi m^2\int\left(\prod_{j=1}^M d\theta_j\right)\frac{K_{\rm fermion}(\theta_1)}{2\theta_1+\pi i} \frac{K_{\rm fermion}(\theta_M)}{2\theta_M-\pi i}\left(\prod_{j=2}^{M-1} K_{\rm boson}(\theta_j)\right)\nonumber\\
&\nonumber\\
&\times\left(\sum_{j=1}^{M} e^{\theta_j}+e^{-\theta_j}\right)^2e^{-2itm\sum_{j=1}^{M}\cosh\theta_j}\nonumber\\
&\nonumber\\
&\times\left\{\left[2\theta_1-\pi i\right]\left[\left(-\theta_1+\theta_2\right)-\pi i\right]\left[\left(-\theta_2+\theta_3\right)-\pi i\right]\times\cdots\times\left[\left(-\theta_{M-1}+\theta_{M}\right)-\pi i\right]\left[2\theta_M+\pi i\right]\right.\nonumber\\
&\nonumber\\
&\times\left.\left[\left(\theta_M-\theta_{M-1}\right)-\pi i\right]\times\cdots\times\left[\left(\theta_3-\theta_2\right)-\pi i\right]\left[\left(\theta_2-\theta_1\right)-\pi i\right]\right\}^{-1}+\mathcal{O}\left(\frac{1}{N}\right),
\eeq
for $M>1$, and
\beq
\frac{1}{N}\langle 0\vert \Theta(t)\vert B_1\rangle=4\pi m^2\int d\theta\frac{K_{\rm boson}(\theta)}{4\theta^2+\pi^2}\left(e^{\theta}+e^{-\theta}\right)^2e^{-2itm\cosh\theta}+\mathcal{O}\left(\frac{1}{N}\right).\nonumber
\eeq

The one-point function is then simply
\beq
\frac{1}{N}\frac{\langle B\vert \Theta(t)\vert B\rangle}{\langle B\vert B\rangle}&=&\frac{\langle \Theta\rangle}{N}+8\pi m^2\int d\theta K_{\rm boson}(\theta) \frac{4\cosh^2\theta \cos(2tm\cosh\theta)}{4\theta^2+\pi^2}\nonumber\\
&&+\sum_{M=2}^\infty4\pi m^2\int\left(\prod_{j=1}^M d\theta_j\right)K_{\rm fermion}(\theta_1) K_{\rm fermion}(\theta_M)\left(\prod_{j=2}^{M-1} K_{\rm boson}(\theta_j)\right)\nonumber\\
&&\nonumber\\
&&\times\frac{2\left(\sum_{j=1}^{M} 2\cosh\theta_j\right)^2 \,\cos\left(2tm\sum_{j=1}^{M}\cosh\theta_j\right)}{\left[4\theta_1^2+\pi^2\right]\left[4\theta_M^2+\pi^2\right]\left\{\prod_{j=1}^{M-1}\left[\left(\theta_j-\theta_{j+1}\right)^2+\pi^2\right]\right\}}+\mathcal{O}\left(\frac{1}{N}\right),\label{solutiontheta}
\eeq
where $\langle \Theta\rangle$ is the vacuum expectation value of $\Theta$, which we can be set to zero, depending on our definition of $\Theta$.

The expression (\ref{solutiontheta}) shows that for long times, the time-dependent one-point function relaxes into  the zero-temperature value. To see this more explicitly, we can write an approximate expression for the ``small quench" limit. This limit corresponds to a case where the post- and pre-quench masses are very close to each other. This implies that $\vert K_{\rm boson}(\theta)\vert,\,\vert K_{\rm fermion}(\theta)\vert<<1$, for all $\theta$. In this limit, we can approximate the function (\ref{solutiontheta}) by keeping only the one Cooper pair contribution (the first term in the right hand side of (\ref{solutiontheta})). For long times, the integral can be computed using the stationary phase approximation. We compute the long-time limit up to order $K(\theta)^2$ (using the explicit functions (\ref{freeboson}) and (\ref{freefermion})), that is, the one and two particle pairs contribution:
\beq
\left.\frac{1}{N}\frac{\langle B\vert \Theta(t)\vert B\rangle}{\langle B\vert B\rangle}\right\vert_{K<<1,\,t\to\infty}&\longrightarrow&\frac{\langle \Theta\rangle}{N} +32 m^2\,\frac{m_0-m}{m_0+m}\,\sqrt{\frac{1}{\pi mt}}\cos\left(2mt+\frac{\pi}{4}\right)\nonumber\\
&&-\frac{128 m^2}{\pi^4}\left(\frac{m_0-m}{m_0+m}\right)^2\left(\frac{1}{m t}\right)\cos\left(4m t+\frac{\pi}{2}\right)+O\left(K^3\right)\label{stationaryphase},
\eeq
with these terms showing a $t^{-1/2}$ and $t^{-1}$ power-law decay, respectively. The series (\ref{solutiontheta}) appears to converge very quickly for small quenches $(m\approx m_0)$, as is seen by plotting the two time-dependent terms in (\ref{stationaryphase}). In Figure 11, we plot separately the $O(K)$ and $O(K^2)$ terms in the stationary phase approximation for long times ($t>\frac{10}{m_0}$) for the values $m_0=1,\,m=1.2$, (The $O(K^2)$ term is so much smaller than the $O(K)$ term that it would be hard to distinguish it if plotted together). 

\begin{figure}
  \centering
    a)\includegraphics[width=0.5\textwidth]{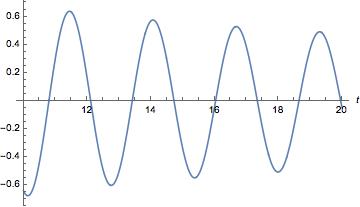}
    
    \vspace{10pt}
    
    \vspace{10pt}
    
    \vspace{10pt}
    
    \vspace{10pt}
    
    b) \includegraphics[width=0.56\textwidth]{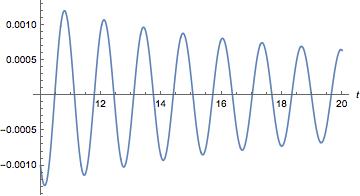}

 \caption{a) The second term in the right hand side of (\ref{stationaryphase}), in units of $m_0^2$, for $m=1.2 \,m_0$. b) The third term in the right hand side of (\ref{stationaryphase}) in the same units. The values in the $y$-axis indicate that b) is already a much smaller contribution, suggesting that the convergence of the series (\ref{solutiontheta}) is very fast.}
\end{figure}

At this point it is very interesting to note that the one-point function does not decay into a thermal expectation value, but into the zero temperature value. The thermal expectation value for $\Theta$ was computed in \cite{thooftthermodynamics} using the Leclair-Mussardo formula \cite{leclairmussardo}. The result from Ref. \cite{thooftthermodynamics} is that there are some nontrivial contributions to the thermal expectation values that are not suppressed by higher powers of $1/N$ compared to the zero-temperature values. In our case, we see that in the long time limit, all of these contributions are gone, and one simple gets the zero temperature values.

The reason why the long-time steady-state observables are given by their zero temperature value is that, as far as the leading $1/N$ contributions are concerned, the averages taken using the generalized Gibbs ensemble corresponding to our color-neutral initial states are equivalent to just the zero-temperature values. The nontrivial contributions to the equilibrium finite temperature values are given by states that are excluded in our quench, by demanding that the initial state is color neutral. Fixing this value of the color charge in the GGE eliminates these particle states that would produce nontrivial thermal contributions. If we perform a quench where the initial state has color-charged contributions, the corresponding GGE can give leading $1/N$ contributions similar to those from Ref. \cite{thooftthermodynamics}.

The fact that the steady state observables are given by their zero temperature averages, instead of by some version of the GGE seems puzzling at first sight. However, as we will discuss further in the conclusions, the $1/N$ suppression of thermal effects is something that can be naturally expected in matrix valued theories. A similar phenomenon in higher dimensional gauge theories is known as Large $N$ volume independence \cite{volumeindependence}. The reason for this suppression is that, in a gauge theory, the finite volume/temperature contributions to expectation values of observables correspond to non-planar Feynman diagrams, and are thus suppressed (as long as the temperature is low enough that center symmetry is not spontaneously broken). The colorless initial states we have studied are center-symmetric invariant, simply because they are color singlets, this means that any thermal-like contributions to the GGE will correspond to non-planar diagrams and will be suppressed.  

We would like to contrast our expression, (\ref{solutiontheta}) with another simple known solution, which is the quantum quench of the Ising model (free fermions). The time-dependent one-point function of the energy operator (which is proportional to $\Theta$) can be computed very easily using a form factor expansion, because only the two-particle form factors are nonzero. The expansion is found in Eq. (47) of \cite{FM}, and has  two  terms, a time independent contribution, which can be described by a GGE, and one decaying time dependent contribution, which is analogous to our one-pair contribution in (\ref{solutiontheta}). In our solution, the time-independent term is suppressed by $1/N$, while the decaying time-dependent part includes an infinite number of terms, due to the fact that the form factors with higher number of particles are nonzero.

\section{Two-point function of the energy-momentum tensor}

We now want to compute the equal-time two point function
\beq
W(x,y;t)=\frac{1}{N^2}\frac{\langle B\vert \Theta (x,t) \Theta(y,t)\vert B\rangle}{\langle B\vert B\rangle}.\label{twopoint}
\eeq
It is very easy to compute this function, again using the fact that the leading $1/N$ contributions come from connected form factors where all the particles form a single string, and that we have chosen $\vert B\rangle$ to be color neutral. 

The way to compute the correlator (\ref{twopoint}) is by inserting a complete set of particle states between the two operators:
\beq
\frac{1}{N^2}\langle B\vert \Theta (x,t) \Theta(y,t)\vert B\rangle=\sum_\Psi \frac{1}{N^2}\langle B\vert \Theta (x,t)\vert \Psi\rangle \langle \Psi\vert \Theta(y,t)\vert B\rangle.\nonumber
\eeq
However, since the states $\vert B\rangle$ are color neutral, all the form factors $\langle B\vert \Theta \vert \Psi\rangle_{\rm connected}$ are suppressed by higher powers of $1/N$, unless $\vert \Psi\rangle=\vert 0\rangle$, because the particles in $\langle B\vert$ and $\vert \Psi\rangle$ cannot form a single closed string.
We can then write the leading contributions to (\ref{twopoint}) as
\beq
W(x,y;t)&=&\sum_{K=0}^\infty\sum_{\sigma^\prime}\sum_{M=0}^\infty \sum_{\sigma} \left(\langle B_K^{(\sigma^\prime)}\vert \Theta(x,t)\vert 0\rangle\right)\left(\langle 0\vert \Theta(y,t)\vert B_M^{(\sigma)}\rangle\right)\nonumber\\
&&+\sum_{\Psi}\langle 0\vert \Theta(x,t)\vert \Psi\rangle\langle\Psi\vert \Theta(y,t)\vert 0\rangle,\label{twopointexpanded}
\eeq
where again, infrared divergences have been canceled with the denominator, $\langle B\vert B\rangle$.

We now substitute the exact form factors in Eq. (\ref{twopointexpanded}) and obtain, at leading order in $1/N$,
\beq
W(x,y;t)&=&W_0(x,y)\nonumber\\
&&+16\pi^2 m^4\left\{\frac{\langle \Theta\rangle}{4\pi m^2N} +\int d\theta K_{\rm boson}(\theta)\frac{4\cosh^2\theta\exp\left(-2itm\cosh\theta\right)}{4\theta^2+\pi^2}\right.\nonumber\\
&&+\sum_{K=2}^\infty \left[\int\left(\prod_{j=1}^K d\theta_j\right)K_{\rm fermion}(\theta_1)K_{\rm fermion}(\theta_K)\left(\prod_{j=1}^{K-1} K_{\rm boson}(\theta_j)\right)\right.\nonumber\\
&&\left.\times\left.\frac{\left(\sum_{j=1}^K2\cosh\theta_j\right)^2\exp\left(-2itm\sum_{j=1}^K \cosh\theta_j\right)}{\left[4\theta_1^2+\pi^2\right]\left[4\theta_K^2+\pi^2\right]\left\{\prod_{j=1}^{K-1}\left[\left(\theta_j-\theta_{j+1}\right)^2+\pi^2\right]\right\}}\right]\right\}\nonumber\\
&&\times\left\{\frac{\langle \Theta\rangle}{4\pi m^2N} +\int d\theta K_{\rm boson}(\theta)\frac{4\cosh^2\theta\exp\left(2itm\cosh\theta\right)}{4\theta^2+\pi^2}\right.\nonumber\\
&&+\sum_{M=2}^{\infty}\left[\int\left(\prod_{j=1}^M d\theta_j\right)K_{\rm fermion}(\theta_1)K_{\rm fermion}(\theta_M)\left(\prod_{j=1}^{M-1} K_{\rm boson}(\theta_j)\right)\right.\nonumber\\
&&\times\left.\left.\frac{\left(\sum_{j=1}^M2\cosh\theta_j\right)^2\exp\left(2itm\sum_{j=1}^M \cosh\theta_j\right)}{\left[4\theta_1^2+\pi^2\right]\left[4\theta_M^2+\pi^2\right]\left\{\prod_{j=1}^{M-1}\left[\left(\theta_j-\theta_{j+1}\right)^2+\pi^2\right]\right\}}\right]\right\},\label{resulttwopoint}
\eeq
where
\beq
W_0(x,y)=\langle 0\vert \Theta(x,t)\Theta(y,t)\vert 0\rangle,\nonumber
\eeq
is the ordinary two point function at equilibrium, which has been found in \cite{correlation}, and we have used the fact that the initial state has zero total momentum. An interesting feature of the expression (\ref{resulttwopoint}), is that the terms with time dependence, and those with dependence on the spatial positions, $x$ and $y$ are separated, such that one can write $W(x,y;t)=W_0(x,y)+W(t)$, where $W(t)$ is defined by Eq. (\ref{resulttwopoint}).

As we did in the previous section for the one-point function, we can also find the long-times and small quench limit of (\ref{resulttwopoint}), using the stationary phase approximation. If we expand $W(t)$ in powers of the functions $K_{\rm boson}(\theta)$ and $K_{\rm fermion}(\theta)$, the leading contribution is
\beq
W(x,y;t)\vert_{K<<1,t\to\infty}\longrightarrow W_0(x,y)+\frac{32 m^2\langle\Theta\rangle}{N}\,\frac{m_0-m}{m_0+m}\,\sqrt{\frac{1}{\pi mt}}\cos\left(2mt+\frac{\pi}{4}\right)+O\left(K^2\right).
\eeq

\section{A comment on correlators of the renormalized field}

We have computed one- and two-point functions of the energy momentum tensor in the previous section for two important reasons, all the form factors of this operator are known at large $N$, and they are nonzero for color-neutral particle states. Another example of an operators whose form factors are known is the renormalized field, $\Phi$. This field is defined in terms of the bare field $U$ by
\beq
\frac{1}{N}\langle 0\vert {\rm Tr} \Phi(x)\Phi(0)^\dag\vert 0\rangle=\frac{1}{N}Z[g_0(\Lambda),\Lambda]^{-1}\langle 0\vert {\rm Tr} U(x) U(0)^\dag\vert 0\rangle,\nonumber
\eeq
where $Z[g_0(\Lambda),\Lambda]$ is a renormalization constant, and $\Lambda$ is an Euclidean momentum cutoff.

The one-point function of the renormalized field operator after a color-neutral quantum quench is trivial, simply because $\Phi$ is not a color-neutral operator. As was shown in \cite{renormalizedfield}, the form factors  of $\Phi$ with excitations in the incoming state are only nonzero for states with $M$ antiparticles and $M+1$ particles, for any integer $M$. The states $\vert B_M^{(\sigma)}\rangle$ have an equal number of particles and antiparticles, we can therefore write
\beq
\langle B\vert \Phi(x,t)\vert B\rangle=0.\nonumber
\eeq

We can then naturally ask the question, is the two-point function of $\Phi$ also trivial? The operator ${\rm Tr}\Phi(x)\Phi(0)^\dag$ is indeed color neutral, so it should have some non vanishing form factors with the states $\vert B_M^{(\sigma)}\rangle$. However, as we will argue, these form factors contribute at subleading orders of $1/N$.

Let us consider the two-point function 
\beq
W^\Phi (x,y;t)=\frac{1}{N}\frac{\langle B\vert {\rm Tr} \Phi(x,t)\Phi(y,t)^\dag\vert B\rangle}{\langle B\vert B\rangle}.\label{fieldtwopoint}
\eeq
We can expand (\ref{fieldtwopoint}) as
\beq
W^\Phi (x,y;t)&=&W^\Phi_0(x,y)\nonumber\\
&&+\frac{1}{N}\sum_{K=0}^\infty\sum_{\sigma^\prime}\sum_{M=0}^\infty \sum_{\sigma} \sum_{\Psi_{\rm odd}}\left(\langle B_K^{(\sigma^\prime)}\vert \Theta(x,t)\vert \Psi_{\rm odd}\rangle\right)\left(\langle \Psi_{\rm odd}\vert \Theta(y,t)\vert B_M^{(\sigma)}\rangle\right)/\langle B\vert B\rangle,\nonumber
\eeq
where $\Psi_{\rm odd}$ are states which have an odd number of excitations, and with $k+1$ particles and $k$ antiparticles, for some integer $k$, and 
\beq
W_0^\Phi(x,y)=\frac{1}{N}\langle 0\vert {\rm Tr}\Phi(x,t)\Phi(y,t)^\dag\vert 0\rangle,\nonumber
\eeq
is the standard equilibrium two-point function, computed in \cite{renormalizedfield}, which has a nontrivial $\mathcal{O}(1/N^0)$ contribution. As we have stated before, the excitations in the states $\vert B_M^{(\sigma)}\rangle$ form a single closed string of color contractions, therefore, any form factors of the forms
\beq
\langle B_K^{(\sigma^\prime)}\,^+\vert \Phi(x,t)\vert \Psi_{\rm odd}\rangle, \,\,\,\,\,\,\,\,\,\,\,\,\,\,\,\,\langle \Psi_{\rm odd}\,^+\vert \Phi(y,t)^\dag\vert B_M^{(\sigma)}\rangle,\nonumber
\eeq
contribute to the correlator (\ref{fieldtwopoint}) only at sub-leading orders in $1/N$, because these are form factors with color contractions forming more than one string, like those pictured in Figure 3.b. We can therefore simply conclude
\beq
W^\Phi(x,y;t)=W_0^\Phi(x,y)+\mathcal{O}\left(\frac{1}{N}\right).\label{finalfield}
\eeq
In other words, the out-of-equilibrium effects of the quantum quench for the renormalized field two-point function are suppressed by higher powers of $1/N$.

\section{Conclusions}

 The study of the out-of-equilibrium dynamics of the PCSM is greatly simplified in the planar limit. In this limit, particles can only interact nontrivially if they share a color index contraction. As a consequence, many form factors vanish, and only certain color combinations in the particle states give nontrivial contributions to correlation functions. 

We were able to identify a simple, but nontrivial class of integrable initial states, only requiring that the state be color neutral, and keeping only the contributions to the initial state which yield non vanishing form factors. These requirements strongly restrict the form of the initial states we consider. We presented a simple diagrammatic way to classify which color-contraction combinations are allowed in these initial states. The particle-pair creation amplitudes in the initial state can be simply expressed in terms of those of a free bosonic and a free fermionic theory. We conjecture that the initial state corresponding to a genuine mass quench in the PCSM may correspond to (or be approximated by) using the known free bosonic and free fermionic amplitudes for a mass quench, which can be determined by a Bogoliubov transformation.

As a consequence of demanding that the initial state be color-neutral, we find that pre-quench particles can be transmitted through the boundary at $t=0$, while keeping full factorization at all times. This means that the planar limit provides a way to work around the negative results from \cite{Delfino,Schuricht}. This is the first known case of a quantum quench that is fully factorizable, yet preserves nontrivial post-quench interactions.

The most important use of the planar limit is to compute exactly out-of-equilibrium correlation functions. All the form factors of several operators of the PCSM are known exactly at large $N$. Using the color-neutral initial states, we can express these correlators as a sum over form factors. These form factor sums are still generally very difficult to compute in integrable field theories, however, given the simple structure of our initial states, most of the terms of this sum contribute only at subleading orders in $1/N$. The surviving terms in the expansion are very simple to compute explicitly.

An interesting result of our computation is that the observables we compute relax into their zero-temperature values at long times. This is shocking at first sight, given the fact that the thermal expectation values of the PCSM at large N are not trivial, and one would expect that the steady state after the quench is not that different from a thermal state. However, this result can be simply understood in terms of the generalized Gibbs ensemble.  Since we are fixing the initial states to have zero color-charge, the steady state is described by an ensemble where this value of the charge is conserved. The nonvanishing contributions to the thermal expectation values come from states with color charge that are not allowed to contribute in our GGE. To leading order in $1/N$, the GGE averages coincide with the zero-temperature expectation values.

This kind of $1/N$ suppression of thermal effects in field theories is not new. In the context of four-dimensional gauge theories, a similar phenomenon is known as large-$N$ volume independence \cite{volumeindependence}. The statement of volume independence for gauge theories is that one can compactify one dimension (compactifying the time direction gives the finite temperature case), using periodic boundary conditions, and as long as there is not a  center-symmetry breaking transition, the finite size effects are suppressed by higher powers of $1/N$. One can also work around this symmetry breaking transition by introducing twisted periodic boundary conditions \cite{volumeindependence}. In our case, the color-neutral initial state for the PCSM behaves in a similar way to the twisted boundary conditions in a gauge theory. 

There are still several open questions we would like to address in the future. First, it is still uncertain whether the color-neutral initial states we consider can correctly describe a mass quench. So far we can only say that our initial states have properties we would expect from a mass quench, which we expect to be integrable and color neutral. A stronger motivation for considering these initial states can come from a computation similar to that of Ref. \cite{Spyros}.

 It would also be very interesting to find $1/N$ corrections to our results. This, however, seems like an extremely difficult task, since there are many steps in our calculation where we neglected $1/N$ corrections, and this is what made our calculation possible. We would first need to find corrections to the form factors, for which only the leading $1/N$ contributions are known. Perhaps the most difficult part of computing any $1/N$ corrections to our results, is that there are infinitely many more terms we need to include in our form factor sums, and these form factors which will have particles in the incoming and outgoing states are generally much harder to compute, since one has to deal carefully with annihilation-pole singularities (as was done for the sine-Gordon model in \cite{Bertin}).
 
We would like to better understand if there is any deeper connection at infinite $N$ between twisted periodic boundary conditions and a quantum quench. Inspired by the result of this paper, we would like to test if an appropriate quantum quench can always be chosen in a higher dimensional gauge theory, such that it prevents the breaking of center symmetry, even for a highly energetic quench. This would lead to a GGE that displays volume independence, as we see in our case.

Studying higher dimensional gauge theories is also interesting because it is known that the planar limit provides a way to work around the Coleman-Mandula theorem. In higher dimensions, a consequence of this theorem is that only free theories can be integrable.   For free theories in higher dimensions, quantum quenches and the GGE have been studied in Ref. \cite{gabrielle}. In the large $N$-limit, however, one can still have nontrivial integrable theories. The best known example is $\mathcal{N}=4$ super Yang-Mills theory, where there is a well established program for computing observables exactly using integrability \cite{beisert}. Another intriguing new example has been proposed in \cite{fluxtube}, where they look for integrable theories that can describe the confining flux-tube worldsheet in four-dimensional, large-$N$ QCD. The planar limit gives some hope to study analytically out-of-equilibrium dynamics in higher dimensions, since one can have nontrivial integrable theories. It would be interesting to understand the relaxation into a steady state of these higher-dimensional models at long times after a quantum quench, and see whether this state can be described by a GGE.

\begin{flushleft}\large
\textbf{Acknowledgements}
\end{flushleft}
I would like to thank Bruno Bertini and Lorenzo Piroli, for carefully reading this manuscript, and for their very useful comments and suggestions. I also thank Gesualdo Delfino for some helpful discussions of the results. This work has been supported by the ERC, under grant number 279391 EDEQS.

\vspace{3mm}



\end{document}